\documentclass[aps.pra,onecolumn,showpacs,preprintnumbers,amsmath,amssymb]{revtex4}
\usepackage{mathrsfs}
\usepackage{graphicx}
\usepackage{amsmath}
\usepackage{ulem}
\usepackage{color}

\begin{document}

\title{Radiation trapping effect versus superradiance in quantum simulation of light-matter interaction}
\author{S. V. Remizov$^{1,2}$, A. A. Zhukov$^{1,3}$, W. V. Pogosov$^{1,4}$, Yu. E. Lozovik$^{1,5,6}$}

\affiliation{$^1$Dukhov Research Institute of Automatics (VNIIA), 127055 Moscow, Russia}
\affiliation{$^2$V. A. Kotel'nikov Institute of Radio Engineering and Electronics, Russian Academy of Sciences, 125009 Moscow, Russia}
\affiliation{$^3$National Research Nuclear University (MEPhI), 115409 Moscow, Russia}
\affiliation{$^4$Institute for Theoretical and Applied Electrodynamics, Russian Academy of
Sciences, 125412 Moscow, Russia}
\affiliation{$^5$Institute of Spectroscopy, Russian Academy of Sciences, 142190 Moscow region,
Troitsk, Russia}
\affiliation{$^6$Moscow Institute of Electronics and Mathematics, National Research University Higher School of Economics, 101000 Moscow, Russia}

\begin{abstract}
We propose a realization of two remarkable effects of Dicke physics in quantum simulation of light-matter many-body interactions with artificial quantum systems. These effects are a superradiant decay of an ensemble of qubits and the opposite radiation trapping effect. We show that both phenomena coexist in the crossover regime of a "moderately bad" single-mode cavity coupled to the qubit subsystem.
Depending on the type of the initial
state and on the presence of multipartite entanglement in it,
the dynamical features can be opposite resulting either in the
superradiance or in the radiation trapping despite of the fact that the initial state contains
the same number of excited qubits. The difference originates from the symmetrical or nonsymmetrical character of the initial wave function of the ensemble, which corresponds to indistinguishable or distinguishable emitters.
We argue that a coexistence of both effects can be used in dynamical quantum simulators to demonstrate realization of Dicke physics, effects of multipartite quantum entanglement, as well as quantum interference and thus to deeply probe quantum nature of these artificial quantum systems.
\end{abstract}


\author{}
\maketitle
\date{\today }

\section{Introduction}

Artificial quantum circuits can be used for the construction of programmable quantum computers of a large scale. Such systems can also serve as a platform for realization and experimental exploration of various fundamental phenomena, which are not easy to observe in the case of natural quantum systems.

One of the most interesting phenomena in the field of quantum optics is Dicke superradiant decay of an ensemble of spins (atoms) interacting with the electromagnetic environment \cite{Dicke}. However, for large number of atoms a direct observation of superradiance is not easy due to the infinite number of modes in the free space, as well as dipole-dipole interaction and diffraction effects \cite{Haroche}. It is therefore of interest to turn to mesoscopic ensembles consisting of relatively small number of artificial atoms (qubits) \cite{DeVoe,Eschner,Fedorov1,Fedorov2} and coupled to the single-mode cavity \cite{Wallraff,coop,superrad,hyperrad}. Moreover, an individual addressability of qubits in such artificial circuits with limited number of qubits can make its possible to create initial states of different types including Dicke states and to bring them to the resonance with the cavity using high flexibility of the circuits. These states, in general case, are characterized by the multiparticle entanglement and they can be engineered, for example, using a standard set of quantum gates, see, e.g., Refs. \cite{Blatt1,Blatt} dealing with the algorithmic preparation of $W$ states involving 8 ion qubits. Although this requires very precise quantum operations, a generation of highly entangled Dicke states for mesoscopic ensembles seems to be realistic for near-term quantum technologies based on trapped ions and superconducting Josephson realizations \cite{Blatt20,supremacy1,Preskill}. For other methods and proposals to generate Dicke states based on the free evolution or projective measurements, see Refs. \cite{Eberly,Haroche,Zanthier}. Alternative physical systems prospective for observation of superradiance are quantum dots \cite{Leymann,Lag}, atoms in optical traps \cite{Optic,atoms}, atomic vapors \cite{Haroche}, NV centers in diamond \cite{NV1}, as well as spins in microwave cavities \cite{spin}.

We here argue that there exists another interesting effect arising from the Dicke physics, which can be referred to as the radiation trapping effect \cite{Cummings}. This effect corresponds to the opposite limit of a qubit-cavity coupling much stronger than cavity dissipation rate and occurs if few particular qubits of the ensemble have been initially excited, so that the initial state contained no entanglement. It is manifested through the increased time scale for the emission of photons to the cavity compared to the single-qubit circuit -- the larger the number of unexcited qubits in the ensemble the longer the emission time and the smaller the photon number in the cavity \cite{Dickedyn}. Thus excited qubits somehow feel the presence of unexcited qubits which block their radiative relaxation to the cavity. On the contrary, in the case of Dicke states, the initial excitations are distributed symmetrically among the qubits in the ensemble in such a way that they are indistinguishable within this collective state. This state contains multipartite entanglement, and the free evolution from this state is superradiant. In both cases, qubits do not behave as independent emitters, but the consequences of this fact are the opposite. In addition, in contrast to the radiation trapping effect, the superradiance can exist in the limit of a bad cavity \cite{superrad}.

In the present article, we show that the dynamical behavior of both types is pronounced and therefore can be observed in the same system, the necessary condition being "moderately bad" cavity coupled to the qubit subsystem. Both effects emerge in this crossover regime, but disappear in opposite limits. We argue that it is of interest to realize these effects in artificial quantum systems which can be treated as dynamical quantum simulators. Such simulators can probe quantum nature of artificial systems on a much deeper level compared to more standard spectroscopic experiments, see, e.g., Refs. \cite{Ustinov1,Ustinov2,Saito,Zagoskin}. The predicted features can be used for a direct demonstration of realization of Dicke physics and effects of multipartite entanglement in such artificial circuits. Note that the radiation trapping effect should not be mixed with the subradiant behavior \cite{Dicke,subrad1,subrad2} occurring when the initial state is entangled, but antisymmetric.

The paper is organized as follows. In Section II we present our model and theoretical tools used to study the dynamics of qubit-cavity system. Section III deals with the system without inhomogeneous broadening as well as with negligibly weak longitudinal and transverse relaxation of qubits. Section IV addresses 'imperfect' systems which are characterized by finite splitting in excitation energies as well as nonzero relaxation rates in qubit subsystem. We conclude in Section V.

\section{Model}

Let us consider a dissipative evolution of an ensemble of qubits coupled to the single-mode quantum resonator. The ensemble is, in general, characterized by some distribution in qubits excitation frequencies. Each qubit as well as the resonator is coupled to its own Markovian bath. The whole system is described by the master equation
\begin{equation}
\partial_t\rho(t)-\Gamma[\rho(t)]=-i[H,\rho(t)],
\label{Lindblad}
\end{equation}
where $\rho(t)$ is a density matrix of the qubit-resonator coupled system. The matrix $\Gamma[\rho]$ depends on rates of energy dissipation in the cavity $\kappa$, in each of the qubits $\gamma$, as well as on the pure dephasing rate $\gamma_{\varphi}$. It is given by $
\Gamma[\rho]=\kappa(a\rho a^{\dagger } -\{ a^{\dagger }a, \rho\}/2)+\sum_{j}\left(\gamma(\sigma_{j,-}\rho\sigma_{j,+} - \{\sigma_{j,+}\sigma_{j,-},\rho\}/2)+ \gamma_{\varphi}(\sigma_{j,z}\rho\sigma_{j,z}-\rho)\right)$, where we assumed for the simplicity that the energy dissipation rates, as well as the pure dephasing rates, are the same for all qubits of the ensemble. The Hamiltonian of the qubits-photon coupled system is of the form
\begin{eqnarray}
H=\sum_{j=1}^L \epsilon_j \sigma_{j}^+ \sigma_{j}^- + \omega a^{\dagger } a + g \sum_{j=1}^L (a^{\dagger }\sigma_{j}^- + a \sigma_{j}^+),
\label{Hamiltonian}
\end{eqnarray}
where $a^{\dagger }$ and $a$ are photon creation and annihilation operators, while $\sigma_{j}^\pm$, $\sigma_{j}^z$ are Pauli operators acting in the space of qubits degrees of freedom. Thus, qubits of the ensemble effectively interact with each other through the
photon degree of freedom. The Hamiltonian (\ref{Hamiltonian})
commutes with the operator $\sum_{j=1}^{N_q} \sigma_{j}^+ \sigma_{j}^- +
a^{\dagger } a$ of the total excitation number, i.e., the number
of excited qubits and photons in the cavity. The Hamiltonian
is based on the rotating wave approximation (RWA), which
neglects counterrotating terms of the form $g(a \sigma_{j}^- +
a^{\dagger } \sigma_{j}^+)$ also appearing in the full expression of
the qubit-cavity interaction operator. These terms do not conserve
an excitation number and they can be safely neglected at $g\ll
\omega$ provided the detuning between the cavity and qubits is not
too large, $|\epsilon_j - \omega| \ll \omega$. Notice that
counterrotating terms however are essential in some special
situation, for instance, under the parametric driving which can give rise to the
dynamical Lamb effect
\cite{paper1,paper2,paper4}. Also note that in absence of dissipation, the system we study can be addressed using an exact solution through Bethe-ansatz technique \cite{Dickedyn,NucPhys2017}.

We hereafter focus on the free evolution starting from the excited system,
which contains $N_{ex}$ excitations created in the qubit subsystem. We consider initial conditions of two kinds. In the first situation, $N_{ex}$ \emph{particular} qubits among $N_q$ qubits are initially excited
\begin{eqnarray}
\Psi_{nonsym} (N_{ex},N_{q})= | \underbrace{\downarrow \ldots \downarrow}\limits_{N_q-N_{ex}} \underbrace{\uparrow \ldots \uparrow}\limits_{N_{ex}} \rangle,
\label{psi1}
\end{eqnarray}
so that qubits of the ensemble are distinguishable within this state, which is separable and contains no entanglement. In the second case (Dicke states), $N_{ex}$ excitations are distributed symmetrically over $N_q$ qubits
\begin{eqnarray}
\Psi_{sym} (N_{ex},N_{q})= \widehat{S}\Psi_{nonsym} (N_{ex},N_{q}),
\label{psi2}
\end{eqnarray}
where $\widehat{S}$ is the symmetrization operator; the qubits in this collective state are indistinguishable. For example, $\Psi_{sym} (1,2)=1/\sqrt(2) (|\uparrow \downarrow \rangle + | \downarrow \uparrow \rangle)$. Dicke states are apparently characterized by the multipartite entanglement \cite{Multipart}. Using individual addressability of qubits in the artificial quantum systems, it might be possible to create also antisymmetric entangled states, which, in the limit of weak cavity dissipation rate, must be weaker coupled to light. The latter corresponds to the so called subradiant behavior \cite{Dicke,Krimer1,subrad2}.

We solve the master equation numerically for the number of
qubits in the ensemble $N_q \lesssim 10$ that corresponds to mesoscopic ensembles. The density matrix has a size $2^{N_{q}+N_{ex}} \times 2^{N_{q}+N_{ex}}$,
so that the numerical solution is not too involved. We assume that $g/ \omega \ll 1$. As long as this condition is satisfied, our main conclusions remain the same. For illustrative purposes, we choose $g/ \omega =0.012$ that is typical, for example, for superconducting transmon-qubits. We also introduce the dimensionless time $\tau$ defined as $\tau = 4 t g^2 /\kappa$, where $\kappa /4 g^2$ is a time scale for the decay of a single qubit to the leaky cavity.

We evaluate the mean number of photons in the cavity $n_{ph}$ and mean total population of upper states of qubit subsystem $n_q$ as a function of $\tau$. The important quantities are $\max\limits_{\tau} n_{ph}(\tau)$ and $\max\limits_{\tau} \dot n_{ph}(\tau)$, which are the maximum number of photons in the cavity and the maximum growth rate of this quantity. We will also use another characteristics to quantify the dynamics, which is the maximum emission rate of qubit subsystem $\max\limits_{\tau} (-\dot n_{q}(\tau))$ taken either as a whole or per the initial excitation $N_{ex}$. For different physical realizations of qubit-cavity coupled systems as well as experimental setups, it may happen that different quantities of this sort are more appropriate for measurement; that is why we consider all of them.

\begin{figure}[h]\center
\includegraphics[width=.95\linewidth]{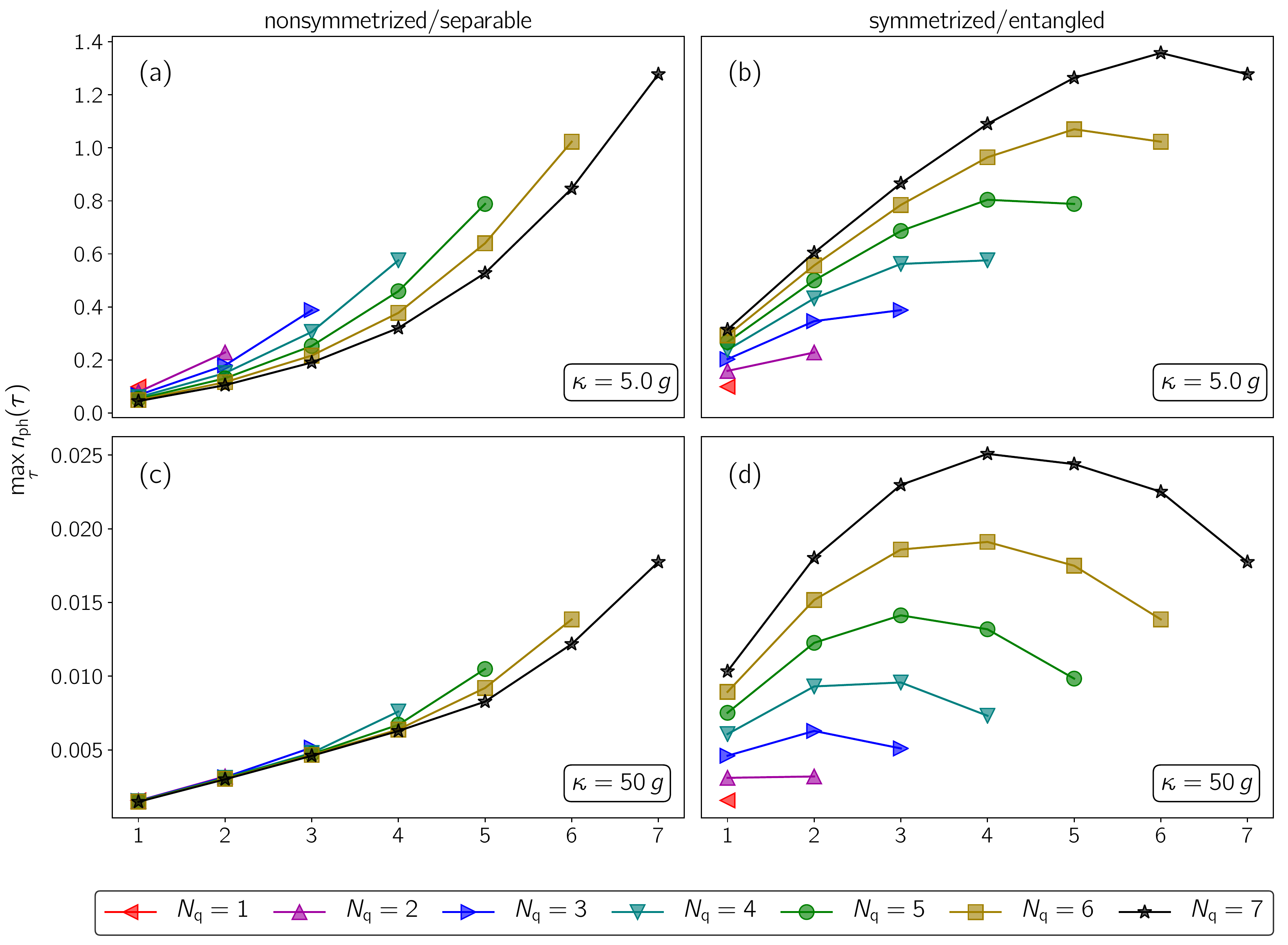}
\caption{Maximum number of photons in the cavity  $\max\limits_{\tau} n_{ph}(\tau)$ achieved during the evolution of the system starting from initial states of two types (\ref{psi1}) and (\ref{psi2}) at $g/ \omega =0.012$.} \label{nph}
\end{figure}

\section{Homogeneous system}

A remarkable effect of Dicke physics is the so-called
radiation trapping effect \cite{Cummings}. It occurs in the limit of weak dissipation, $g \gg \kappa, \gamma, \gamma_{\phi}$ (the ideal platform to observe it is a closed quantum system),  provided there
are many identical two-level systems (qubits) interacting resonantly
with the single-mode radiation field, while a small fraction of them is initially
excited in such a way that they become distinguishable. This corresponds precisely to the initial state of the form (\ref{psi1}). According to this scenario, the larger $N_q$ at fixed $N_{ex}$ the less photon number $\max\limits_{\tau} n_{ph}(\tau)$ is released. The same applies for both $\max\limits_{\tau} \dot n_{ph}(\tau)$ and $N_{ex}^{-1}\max\limits_{\tau} (-\dot n_{q}(\tau))$. In other words,
 the presence of the environment of remaining qubits, though
they are in their ground states, strongly affects dynamics of
the particular excited qubits by blocking their radiative relaxation to
the cavity.

\begin{figure}[h]\center
\includegraphics[width=.95\linewidth]{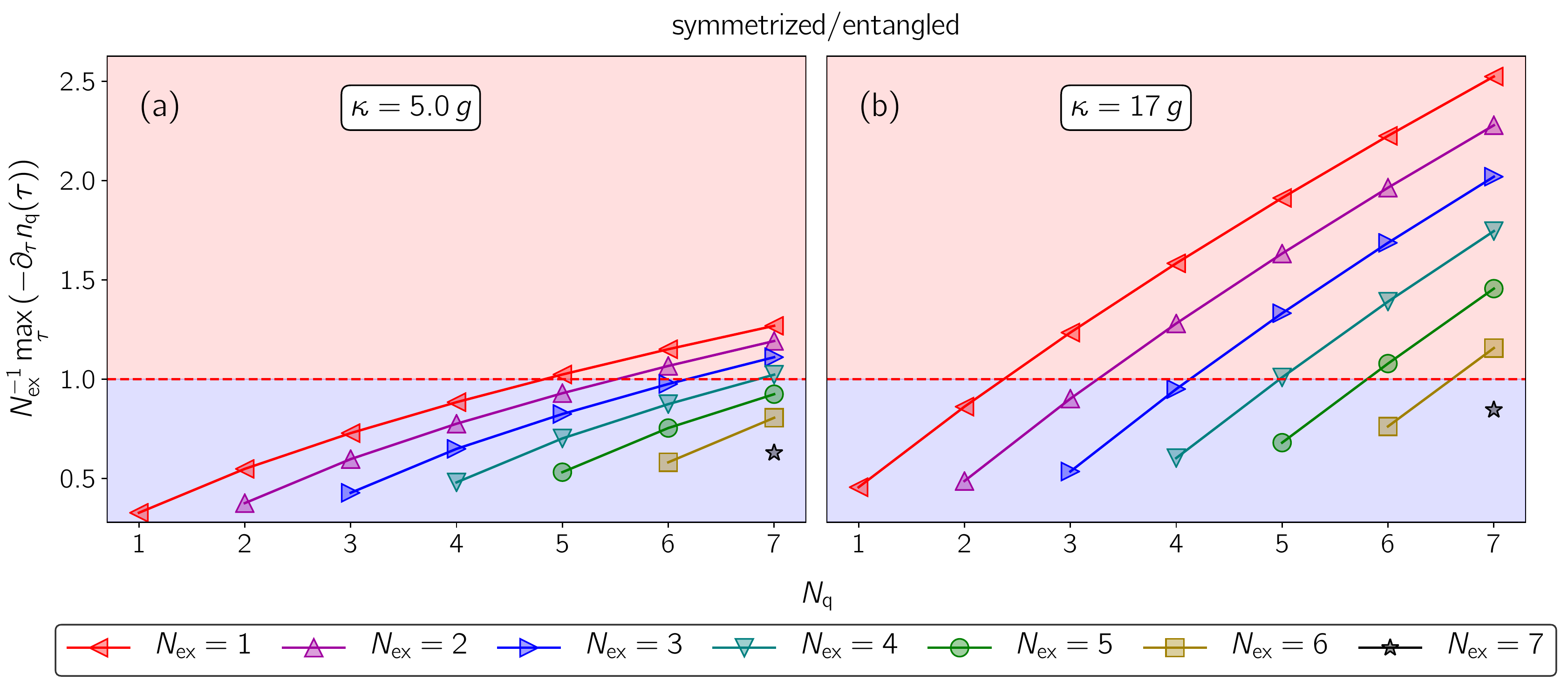}
\caption{Maximum emission rate of qubit subsystem per initial excitation $N_{ex}^{-1}\max\limits_{\tau} (-\dot n_{q}(\tau))$ as a function of qubit number $N_q$ for different values of $\kappa / g$ at $g/ \omega =0.012$. Initial states of the system are Dicke states (\ref{psi2}).} \label{sigmarate}
\end{figure}

In the opposite limit of a bad single-mode cavity, $\kappa \gg g$, the states of the second type (\ref{psi2}), which are Dicke states, show different behavior - the maximum emission rate  $N_{ex}^{-1}\max\limits_{\tau} (- \dot n_{q}(\tau))$ is increased as $N_q$ increases (at fixed $N_{ex}$). Moreover, $\max\limits_{\tau} (- \dot n_{q}(\tau))$ maximized also with respect to $N_{ex}$, i.e., $\max\limits_{N_{ex}}\max\limits_{\tau} (- \dot n_{q}(\tau))$, grows quadratically as a function of $N_q$, which is usually considered as one of the crucial signatures of a superradiance. This maximum is attained at $N_{ex} \approx N_q / 2$ and it provides a highest possible emission rate from the ensemble with a given number $N_q$ of qubits. Artificial quantum circuits are prospective for the observation of both effects. Particularly, individual addressability of qubits can be utilized to engineer various initial conditions, while the interaction can be embedded into the system on physical level.

In the present Section, we study both the radiation trapping effect and superradiance in presence of energy dissipation in qubit and cavity subsystems, but for all qubits having the same excitation frequencies (homogeneous system). Since qubits can feel each other
only through the photon degrees of freedom, one would naively expect
that the radiation trapping effect must be fragile with respect to
the energy dissipation in the cavity and it should disappear as $\kappa$ approaches $g$. We, however, show that although the radiation trapping effect is indeed fully suppressed in the limit of a very bad cavity, $\kappa \gg g$, it survives and even remains pronounced in the regime of "moderately bad" cavity, $\kappa \gtrsim g$. Interestingly, superradiant effects also emerge in this intermediate regime between the limits of weak and strong coupling. Thus, the systems with parameters falling in this range are prospective for the observation of both effects. The realization of them in artificial quantum systems build from superconducting quantum circuits can serve as a demonstration of Dicke physics.

We begin our analysis from the limit, when energy dissipation as well as pure dephasing in qubit subsystem are negligibly weak. We also postpone the discussion of the effects arising from inhomogeneous broadening, i.e., the difference in $\epsilon$'s, to the next Section.

Figure \ref{nph} shows our results for $\max\limits_{\tau} n_{ph}(\tau)$ achieved during the evolution from the initial states of the two kinds described above. Notice that in the bad cavity limit most of the photons escape the cavity before they are again re-absorbed by qubits. The photon number in the cavity is a directly measurable quantity within different realizations of coupled qubit-cavity systems. Two opposite situations are addressed in Fig. \ref{nph}, which correspond to the case of $\kappa \gtrsim g$ (a, b) and $\kappa \gg g$ (c, d). It is seen from these plots that, for nonsymmetrical initial states (\ref{psi1}), the larger $N_q$ at fixed $N_{exc}$ the smaller $\max\limits_{\tau} n_{ph}(\tau)$. This is a signature of the radiation trapping effect. The comparison between Fig. \ref{nph} (a) and (c) shows that this effect becomes suppressed in the limit of a very bad cavity. On the contrary, the Dicke states (\ref{psi2}) give rise to the opposite behavior -- $\max\limits_{\tau} n_{ph}(\tau)$ is rising as $N_q$ grows (at fixed $N_{exc}$). Moreover, the comparison between Fig. \ref{nph} (b) and (d) evidences that, in the bad cavity limit, the highest photon number at fixed $N_q$ is achieved for $N_{ex} \approx N_q /2$, which is a typical feature of superradiant behavior. As a result, the maximum emission rate from the qubit subsystem scales quadratically as $N_q$ grows. In general, we see that superradiance and radiation trapping effect coexist in the regime of "moderately bad" cavity, $\kappa \gtrsim g$.

\begin{figure}[h]\center
\includegraphics[width=.95\linewidth]{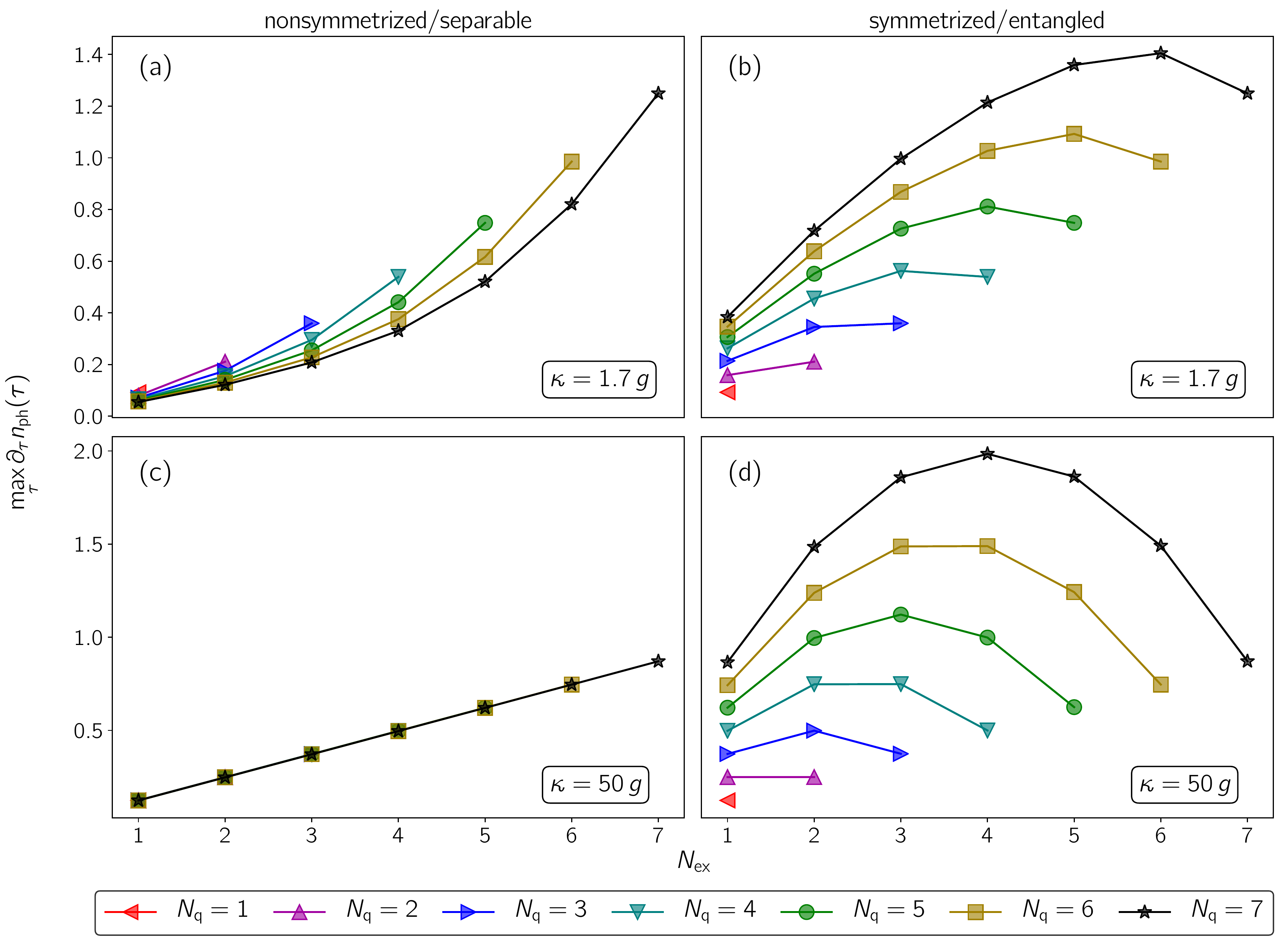}
\caption{Maximum rate of increase of the photon number in the cavity $\max\limits_{\tau} \dot n_{ph}(\tau)$ for initial states of two types (\ref{psi1}) and (\ref{psi2}) at $g/ \omega =0.012$.} \label{nprate}
\end{figure}

Let us now consider $\max\limits_{\tau}  (- \dot n_{q}(\tau))$, which also gives an important information on the superradiant dynamics. It determines the total emission rate for photons which either stay in the leaky cavity or leave it. In experiments, this quantity can be extracted by performing measurements of populations of upper levels. We found that the dependencies of this quantity on $N_{q}$ and $N_{ex}$ are very similar to the dependencies of $\max\limits_{\tau} n_{ph}(\tau)$ as a function of the same parameters, shown in Fig. \ref{nph} (up to the rescaling along the vertical axis). Therefore, we do not present them here. Instead, in Fig. \ref{sigmarate} we show $(N_{ex})^{-1}\max\limits_{\tau} (-\dot n_{q}(\tau))$ as a function of $N_{ex}$ for initial states being (\ref{psi2}). The superradiant behavior occurs provided this quantity is larger than 1, which corresponds to the emission from independent emitters. From this figure it is indeed seen that superradiance is established already at $\kappa$ several times larger than $g$, and the width of corresponding region on the "phase diagram" in $(N_{q},N_{ex})$ plane rapidly grows, as $\kappa / g$ decreases. We remind that the radiation trapping effect does exist in the intermediate range of $\kappa$'s and thus we provide an additional support to our idea that both effects can be observed in the same system.

\begin{figure}[h]\center
\includegraphics[width=.95\linewidth]{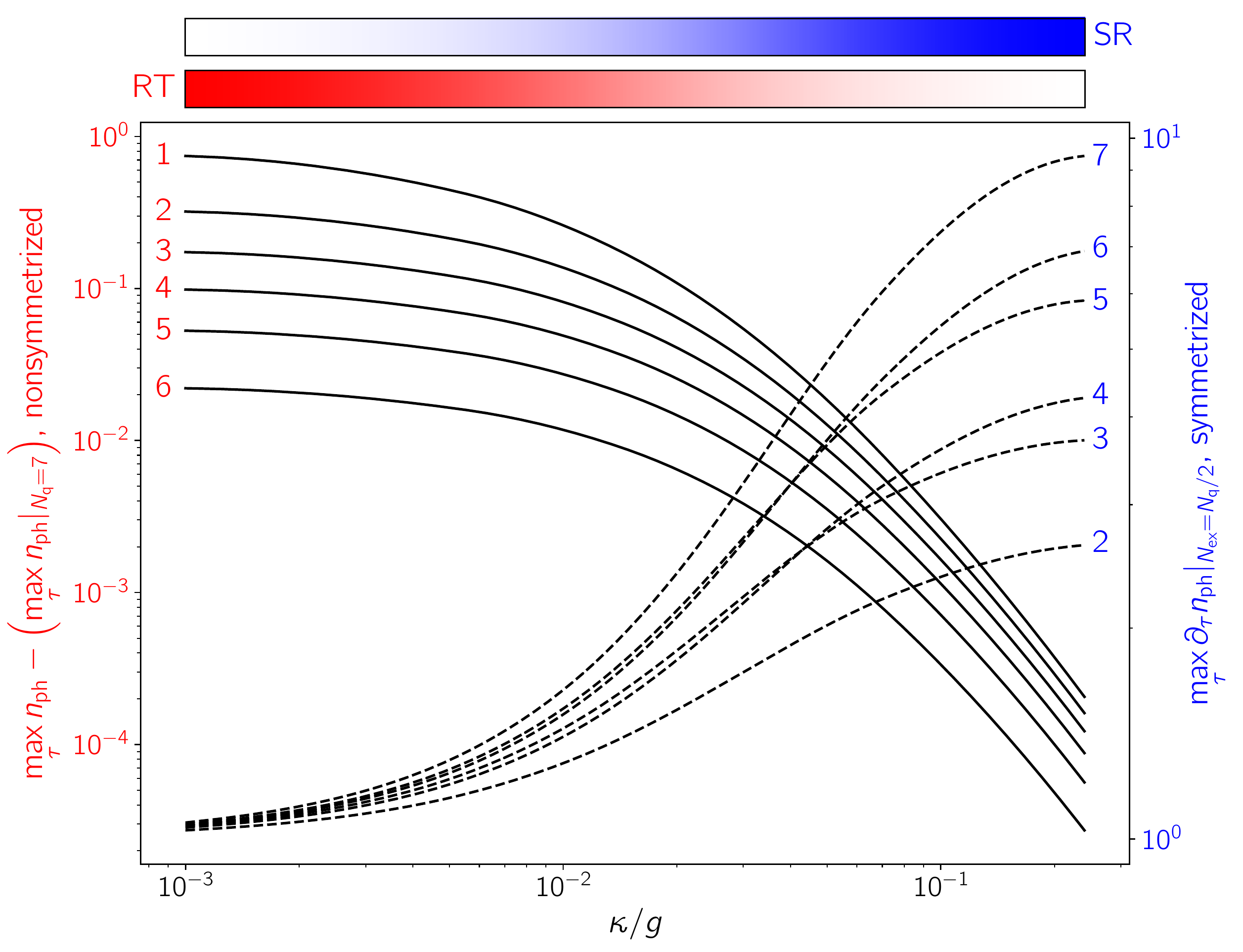}
\caption{The crossover between the radiation trapping (RT) effect at $\kappa \ll g$ and superradiant (SR) behavior at $\kappa \gg g$; $g/ \omega =0.012$, numbers correspond to the numbers of qubits interacting with the photon field (see in the text).} \label{summary}
\end{figure}

We also computed $\max\limits_{\tau} \dot n_{ph}(\tau)$, which may be experimentally relevant provided photon number is measured that depends on a particular physical realization. The results for initial states of both types are presented in Fig. \ref{nprate}. These plots again generally resemble the plots for $\max\limits_{\tau} n_{ph}(\tau)$ shown in Fig. \ref{nph} as well as analogous dependencies for $\max\limits_{\tau} (-\dot n_{q}(\tau))$ (up to the rescaling along the vertical axis due to the finiteness of $\kappa$). The only exception is the limit of a bad cavity and the initial state of the form (\ref{psi1}), which shows no dependence on $N_q$, as seen from Fig. \ref{nprate} (c).


Our general results are illustrated in Fig. \ref{summary}, which deals with the crossover between the two regimes as $\kappa / g$ grows. At small values of this quantity, a radiation trapping effect is realized. It is visualized by the comparison between $\max\limits_{\tau} n_{ph}(\tau)$ at $N_q=7$ and at $N_q < 7$. As $\kappa / g$ is increased, the effect gradually disappears. On the contrary, the superradiant behavior is realized at $\kappa / g \gg 1$. It is visualized through $\max\limits_{\tau} \dot n_{ph}(\tau)$ at $N_{ex}=N_q/2$ for $N_q$ even and at $N_{ex}=(N_q+1)/2$ for $N_q$ odd. This effect disappears at $\kappa \lesssim g$. Both radiation trapping effect and superradiance do coexist in the crossover region $\kappa \sim g$.

\section{Imperfect systems}

It is of interest to explore an influence of disorder in excitation frequencies of individual qubits on both effects we consider. Particularly, it is of importance to understand under what requirements for the broadening the two effects still coexist. Note that it is very difficult to avoid a disorder in excitation frequencies of solid-state qubits due to the limitations of microfabrication technologies. Moreover, finite splitting of frequencies might be necessary in order to achieve individual addressability of qubits in the fixed-frequency architectures. An effective splitting of frequencies can be also induced by a direct dipole-dipole interaction between qubits. We will concentrate on the random distribution of qubit frequencies confined between two cutoffs, the difference between them being $\Omega$, which is a simple, but physically meaningful model.

\begin{figure}[h]\center
\includegraphics[width=.95\linewidth]{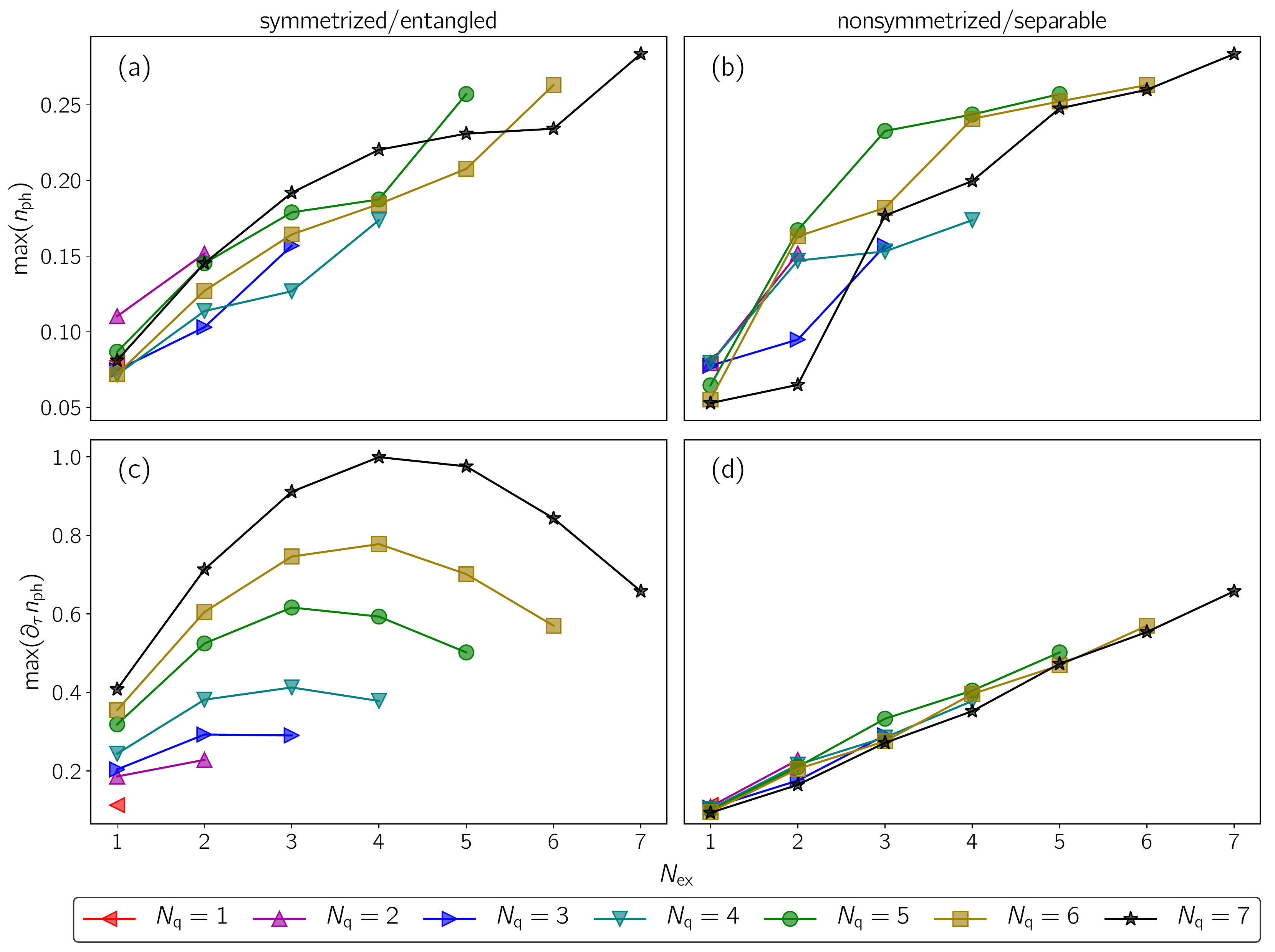}
\caption{The dependencies of $\max\limits_{\tau} n_{ph}(\tau)$ (a, b) and $\max\limits_{\tau} \dot n_{ph}(\tau)$ (c, d) for the initial states of two types -  Dicke states (a, c) and distinguishable qubit states (b, d) at $g/ \omega =0.012$, $\Omega = 0.3 \omega$.} \label{disorder}
\end{figure}

From rather general considerations, it can be expected that the superradiant behavior survives until the effective dephasing time induced by inhomogeneous broadening, $\sim 2\pi / \Omega$, is larger than the duration of the initial superradiant pulse, $\sim   \kappa / g^2$. Thus, superradiant behavior should survive up to $\Omega \lesssim 2\pi g^2/\kappa$. However it is not evident how the radiation trapping  is affected by disorder in the excitation energies of qubits. In order to understand this, we performed numerical computations at various values of controlling parameters including $\Omega$. Figure \ref{disorder} shows typical representative results but for a particular realization of disorder at rather rather large value of $\Omega$ exceeding $2\pi g^2/\kappa$ several times. We indeed found that the above conclusion for the superradiance is valid. We also revealed that, for the initial state being one of the Dicke states, the inhomogeneous broadening changes the characteristic behavior of $\max\limits_{\tau} n_{ph}(\tau)$ much stronger compared to both $\max\limits_{\tau} \dot n_{ph}(\tau)$  and $\max\limits_{\tau} (-\dot n_{q}(\tau))$. Remarkably, the radiation trapping effect also survives up to this range of $\Omega$ that means that both effects do coexist and therefore can be observed in the same system. Note that the results shown in Fig. \ref{disorder} correspond to the larger value of $\Omega$, for which radiation trapping is already suppressed.

We also explored the sensitivity of both phenomena -- radiation trapping effect and superradiance -- to the finite relaxation rates (both longitudinal and transverse) of the qubits. We found that these effects generally survive up to the values of $\gamma$ and $\gamma_{\phi}$ approaching $g^2 / \kappa$, which is an expectable result for the superradiance but not obvious for the radiation trapping effect. As an example, let us consider superconducting qubits. For $\kappa$ exceeding $g$ by a factor of nearly 5, which corresponds to the most appropriate regime for the coexistence of both effects, $T_1$ and $T_2$ must be above nearly 1 microsecond in order to make observations of both effects possible. Of course, such a requirement is readily satisfied in the state-of-the-art superconducting quantum circuits. We also would like to mention that the radiation trapping effect is slightly more sensitive to the finite relaxation rates in the qubit subsystem. In this case, finite $\gamma$ and $\gamma_{\phi}$ suppress characteristic dependencies of both $\max\limits_{\tau} \dot n_{ph}(\tau)$ and $\max\limits_{\tau} \dot n_{q}(\tau)$ on $N_q$ more significantly than similar dependencies of $\max\limits_{\tau} n_{ph}(\tau)$. For the superradiance, the same is true for the dependencies of $\max\limits_{\tau} \dot n_{q}(\tau)$ on $N_q$. However, all features discussed above for the qubits in absence of relaxation remain essentially unchanged if the condition $\gamma$, $\gamma_{\phi} \ll g^2 / \kappa$ is satisfied.

\section{Conclusions}

In the present paper, we argued that the dynamics of the ensemble of qubits coupled to the single-mode cavity can be drastically different depending on the distinguishability of qubits as emitters at the initial moment of time even if the number of excitations stored in the qubit subsystem is the same. This effect exists only in the regime of the "moderately bad" cavity. A superradiant behavior is expected for the initial states with excitations distributed symmetrically over the qubit ensemble (indistinguishable ensemble), which thus contains multipartite entanglement. The opposite behavior resulting in the radiation trapping must be realized if some particular qubits were initially excited, so that the state was not entangled. In both cases, qubits do not behave as independent emitters, but consequences are the opposite. Let us stress that these two effects are maximum in different limits of ratio between cavity dissipation rate and qubit-cavity coupling strength, but they coexist in the crossover regime of a "moderately bad" cavity which thus is most appropriate for the observation of predicted features.

We pointed out that it is prospective to realize both effects in artificial quantum systems which can be considered as quantum simulators of Dicke model and to reveal the impact of multipartite entanglement and quantum interference on their dynamics. Such experiments can provide a deep probe of the "quantumness" of these artificial systems.

The coexistence of both effects was revealed by the numerical solution of master equation for the mesoscopic ensembles of qubits that are most relevant for the state-of-the-art experimental situations. In our studies we concentrated on different types of initial conditions, as well as on the influence of inhomogeneous broadening in the ensemble of qubits. We also analyzed role played by the qubit relaxation rates. As a result, we revealed ranges of main controlling parameters most appropriate for the observation of our predictions. In our studies, we also considered different quantities which can be measured in experiments to detect superradiant behavior and radiation trapping effect. Some of them can be preferable depending on the physical realization of coupled qubit-cavity systems as well as on experimental setups.

\begin{acknowledgments}
Useful discussions with A. V. Ustinov, I. S. Besedin, and E. Andrianov are acknowledged. W. V. P. acknowledges a support from RFBR (project no. 19-02-00421).  Yu. E. L. acknowledges a support from RFBR (project no. 17-02-01134) and from the Program of Basic Research of HSE.
\end{acknowledgments}

\end{document}